%% file: main.tex
\documentclass{sigchi-ext}
\usepackage[T1]{fontenc}
\usepackage{textcomp}
\usepackage[scaled=.92]{helvet} 
\usepackage{graphicx} 
\usepackage{balance}  
\usepackage{booktabs} 
\usepackage{ccicons}  
\usepackage{ragged2e} 
\usepackage{makecell}
\usepackage{adjustbox}
\usepackage{multirow}

\usepackage{xcolor}
\definecolor{pink}{rgb}{0.9,0,0.9}

\def\plaintitle{How Do the Open Source Communities Address Usability and UX Issues? An Exploratory Study} 
\def\emptyauthor{}
\def\plainkeywords{Usability; user experience; open source software development; open source community; issue tracking.}

\title{How Do the Open Source Communities Address Usability and UX Issues? An Exploratory Study}

\numberofauthors{2}
\author{%
    \alignauthor{%
        \textbf{Jinghui Cheng}\\
        \affaddr{Polytechnique Montreal} \\
        \affaddr{Montreal, QC, Canada} \\
        \email{jinghui.cheng@polymtl.ca}
    }
    \alignauthor{} \vfil
    \alignauthor{%
        \textbf{Jin L.C. Guo}\\
        \affaddr{McGill University}\\
        \affaddr{Montreal, QC, Canada}\\
        \email{jguo@cs.mcgill.ca}
    }
    \alignauthor{} \vfil
    \alignauthor{}
    \alignauthor{} \vfil
    \alignauthor{}
}

\definecolor{linkColor}{RGB}{6,125,233}
\hypersetup{%
  pdftitle={\plaintitle},
  pdfauthor={\emptyauthor},
  pdfkeywords={\plainkeywords},
  bookmarksnumbered,
  pdfstartview={FitH},
  colorlinks,
  citecolor=black,
  filecolor=black,
  linkcolor=black,
  urlcolor=linkColor,
  breaklinks=true,
}

\begin{document}

\CopyrightYear{2018}
\setcopyright{rightsretained}
\conferenceinfo{CHI'18 Extended Abstracts}{April 21--26, 2018, Montreal, QC, Canada}
\isbn{978-1-4503-5621-3/18/04} \doi{https://doi.org/10.1145/3170427.3188467}
\copyrightinfo{\acmcopyright}

\maketitle

\RaggedRight{}

\begin{abstract}
    Usability and user experience (UX) issues are often not well emphasized and addressed in open source software (OSS) development. There is an imperative need for supporting OSS communities to collaboratively identify, understand, and fix UX design issues in a distributed environment. In this paper, we provide an initial step towards this effort and report on an exploratory study that investigated how the OSS communities currently reported, discussed, negotiated, and eventually addressed usability and UX issues. We conducted in-depth qualitative analysis of selected issue tracking threads from three OSS projects hosted on GitHub. Our findings indicated that discussions about usability and UX issues in OSS communities were largely influenced by the personal opinions and experiences of the participants. Moreover, the characteristics of the community may have greatly affected the focus of such discussion.
\end{abstract}



\input{s_introduction}
\input{s_methods}
\input{s_findings}
\input{s_discussions}

\balance{}

\bibliographystyle{SIGCHI-Reference-Format}

\end{document}

%% file: s_introduction.tex
\section{Introduction}

Usability and user experience (UX) are software attributes that determine how easy, efficient, error-preventing, and pleasant a software system to be used by human users. Their significance is increasingly pervasive in modern software systems. Establishing good usability and UX often requires a serious commitment from the development team, as well as skilled personnel who possess the required expertise (e.g. UX designers, user researchers). In open source software (OSS) development, however, the community surrounding a software project usually do not have such resource to properly address usability and UX issues \cite{Andreasen2006, Nichols2006}. This is a key reason why many OSS projects suffer from poor user adoption \cite{LisowskaMasson2017}.

Apart from limited resource and experience, the literature has identified several other factors associated with OSS communities that may have resulted in their lack of concentration on these user-centered software qualities. For example, the merits and contribution of good usability and UX practices are not necessarily valued by participants of OSS communities, who in turn, value the quality of source code and feature contributions \cite{Nichols2006, Terry2010}. Consequently, power differences and political conflicts among developers, UX practitioners, and users of an OSS are usually evident \cite{Rajanen2015}. Further, there is currently very little theories, methods, and tools to support OSS communities to collaboratively identify, understand, and address UX design issues in a distributed setting \cite{Raza2012}. As the OSS practice becomes more mature, investigating approaches to facilitate a distributed collaborative environment among developers, UX practitioners, and users has become extremely imperative.

\input{margin_p2}

As a first step towards this effort, we argue that it is important to understand how usability and UX issues are currently being raised, discussed, negotiated, and eventually addressed by the OSS communities. Knowledge in this respect is very limited in the literature. In this paper, we bridge this gap and report on an exploratory study that investigated patterns and themes in selected issue tracking threads
that focused on usability and UX topics from three OSS projects hosted on GitHub.

\subsection{Background: GitHub Issue Tracking System}
Most software projects use a tracking system for reporting, discussing, and managing tasks, enhancements, and bugs of the project. Such systems are also a direct channel for communication among the software developers and users. This role of the tracking systems is especially important for OSS projects as their developers and users are often geographically distributed. In this paper, we focus on the GitHub tracking system named \textit{Issues}. Each issue of a GitHub project is comprised of a title, a description, and a list of comments provided by authorized users. Each issue also possesses a state of either ``open'' or ``closed''. The closed issues are the ones that are resolved and verified by the developers. GitHub also provide features to filter and categorize issues using labels and milestones. 

%% file: margin_p2.tex
\begin{margintable}[1pc]
  \vspace{140pt}
  \begin{minipage}{\marginparwidth}
    \begin{tabular}{l c c}
        \textbf{\thead{Project}}
        & \textbf{\thead{Release\\Year}}
        & \textbf{\thead{Total\\Issues}} \\
        \toprule
        \small{Atom} & \small{2015} & \small{12365} \\
        \small{Eclipse Che} & \small{2016} & \small{3900} \\
        \small{OpenToonz} & \small{2016} & \small{860} \\
        \bottomrule
    \end{tabular}
    
    \vspace{8pt}
    \small{
        \textbf{Atom} is a text and code editor developed by GitHub, with strengths in its support for various plugins.\\
        \vspace{3pt}
        \textbf{Eclipse Che} is a cloud-based IDE developed by the Eclipse Foundation.\\
        \vspace{3pt}
        \textbf{OpenToonz} is a 2D animation software developed by Dwango and originally used by Studio Ghibli.
    }
    \caption{Information about the three OSS projects we chose to analyze.}
    \label{tab:projectInfo}
  \end{minipage}
\end{margintable}

%% file: s_methods.tex
\section{Methods}
In this section, we describe how we chose the OSS projects and threads, as well as our data analysis methods.

In order to understand the current trends in managing usability and UX issues, we focused on popular and recently active OSS projects. Particularly, we chose three recently released projects from the top ten OSS projects picked by opensource.com in 2016 \cite{Huger2016}: Atom, Eclipse Che, and OpenToonz; see Table \ref{tab:projectInfo} for information of each project. We also particularly chose these projects because they targeted different user groups (i.e. general users with text and code editing needs, hard core programmers, and animation artists, respectively). This distinction allowed us to explore the characteristics of different OSS communities.

We utilized the GitHub REST API to extract the issue threads. For each project, we first queried ``closed'' issues whose comments included one of the following keywords: usability, UX, and "user experience" (exact phase). We then sorted these issues based on the number of comments it possessed. Next, one researcher read the issue descriptions and manually selected three issues per project from the top of the list that were truly related to usability and UX. Selecting the most commented threads allowed us to focus on issues that included the richest information about (1) the communities' perceptions on usability and UX and (2) the issues' life cycle. Finally, we downloaded all the comments on each selected issue (see Table \ref{tab:threadsInfo}). All comments were extracted on November 10, 2017.

\begin{table}[ht]
    \centering
    \begin{tabular}{l c l c}
        \toprule
        \textbf{\footnotesize{Project}}
        & \textbf{\footnotesize{ID}}
        & \textbf{\footnotesize{Issue Name}}
        & \textbf{\footnotesize{Comments}} \\
        \midrule
        
        \multirow{3}{*}{\footnotesize{Atom}}
        &\footnotesize{5344} & \footnotesize{\makecell[l]{Tab switching should be in MRU\\ order}} & \footnotesize{149} \\
        &\footnotesize{1722} & \footnotesize{\makecell[l]{Open file in current window}} & \footnotesize{148} \\
        &\footnotesize{307} & \footnotesize{\makecell[l]{Large file support}} & \footnotesize{140} \\
        \midrule
        
        \multirow{3}{*}{\footnotesize{\makecell{Eclipse\\Che}}}
        &\footnotesize{5335} & \footnotesize{\makecell[l]{Show current git branch in the IDE}} & \footnotesize{36} \\
        &\footnotesize{3614} & \footnotesize{\makecell[l]{Improve git commit window}} & \footnotesize{30} \\
        &\footnotesize{5484} & \footnotesize{\makecell[l]{Deprecate and remove subversion\\ support}} & \footnotesize{25} \\
        \midrule
        
        \multirow{3}{*}{\footnotesize{\makecell{Open\\Toonz}}}
        &\footnotesize{1316} & \footnotesize{\makecell[l]{Updated column header layouts}} & \footnotesize{134} \\
        &\footnotesize{417} & \footnotesize{\makecell[l]{libmypaint for brush tool in raster\\ levels}} & \footnotesize{64} \\
        &\footnotesize{102} & \footnotesize{\makecell[l]{Ability to parent things to mesh\\ bones}} & \footnotesize{45} \\
        \bottomrule
    \end{tabular}
    \vspace{-4pt}
    \caption{Information about the issue threads we chose to analyze}
    \label{tab:threadsInfo}
    \vspace{-8pt}
\end{table}

We conducted an inductive qualitative analysis to identify themes in the issue discussions. The two researchers first independently coded the threads \cite{saldana_coding_2009}. We then discussed and reached an consensus for the themes to code and created a codebook. One researcher then refined the coding based on the codebook. In the following sections, we report the themes that we identified and our findings on the differences among the three OSS communities.

%% file: s_findings.tex
\section{Themes in Issue Discussion}
The themes we identified were arranged based on the life cycle of the usability/UX issues, including (1) issue reporting, (2) current state, (3) desired state, (4) commitment, and (5) fix/implementation.

\subsection{Issue Reporting}
When the thread participants reported (i.e. created) the usability/UX issues, they discussed justifications for the report that we categorized into three groups:

\vspace{-4px}
\textbf{Personal usage experience} (coded in four issues). E.g., the creator of Atom 307 wrote: ``\textit{One of the things I do most often ... is open up large log files. ... I just tried viewing a 350MB log file and Atom locked up immediately.}''

\vspace{-4px}
\textbf{Experience of a competitor} (coded in four issues). E.g., in Atom 1722, the creator cited another user's comment: ``\textit{I'm coming from Sublime Text 2 and the default behavior there is every file in a new window.}''

\vspace{-4px}
\textbf{Personal opinion} (coded in five issues). E.g., the creator of Eclipse Che 3614 wrote: ``\textit{Git commit window needs to be improved.}''

\subsection{Current State}
Once the issue was created, the thread participants usually started by discussing the current state of the software associated with the issue (i.e. the problem space). We identified three themes in these discussions:

\vspace{-4px}
\textbf{Urgency and/or severity} (coded in 117 comments; \textit{n}=117). Very often, participants voiced needs to change the current state of the software by, for example, claiming that the current state is problematic or mentioning that they have to use a competitor or workaround. E.g., in Atom 307, one participant wrote: ``\textit{This is a HUGE problem}'' and ``\textit{this limitation makes it pretty useless for day to day work.}'' On the other hand, participants sometimes argue for the opposite, claiming it is not necessary or not urgent to address the issue. E.g., in Eclipse Che 5484, one participant was ``\textit{not sure it's time to remove (SVN support)}'' because ``\textit{it seems there are enterprises which still support only SVN.}''

\vspace{-4px}
\textbf{Symptoms caused by the issue} (\textit{n}=55). Thread participants also often describe how the issue had affected their experience using the software. For example, in Eclipse Che 5335, one participant wrote: ``\textit{When I use Che day to day, I have several git projects in my workspace and it will be confusing to have one branch name in the bottom which I don't know to which project it refers.}''

\vspace{-4px}
\textbf{Complexity and/or difficulty} (\textit{n}=19). Sometimes, participants described complexity of the issue and/or difficulty fixing the issue because of historical matters or dependencies on other parts of the software. For example, in Atom 307, a developer wrote: ``\textit{We made a pragmatic decision early on to perform editor state manipulations synchronously for a more convenient scripting experience, but we may need to revisit that decision.}''

\subsection{Desired State}
Throughout the threads, participants discuss the changes they want to make or the design they think could resolve the issue (i.e. the solution space). We categorized these discussions into four groups.

\vspace{-4px}
\textbf{Tentative design/solution} (\textit{n}=35), in which participants provided initial design ideas that could resolve the issue. E.g., when the thread participant created Eclipse Che 3614, he or she also provided a design mock-up specifying how the participant thought the UI could be improved.

\vspace{-4px}
\textbf{Feedback and responses} (\textit{n}=187), in which participants provided suggestions for improving or augmenting the proposed design/solution. E.g., in Eclipse Che 3614, at least five participants joined the discussion about improving the initially proposed design.

\vspace{-4px}
\textbf{Clarifications} (\textit{n}=43), in which thread participants seek or provide clarifications on details of the proposed design or feedback. E.g., in Eclipse Che 5335, a user asked: ``\textit{@slemeur If the footer bar at the bottom of the IDE is gone with Che 6, where current widgets from that panel (maven loader) will be placed?}'' Then slemeur replied: ``\textit{The proposal we choose is to use a toast notification which will be displayed in bottom right corner...}''

\vspace{-4px}
\textbf{Learning from another software} (\textit{n}=34), in which participants pointed out that it worth learning lessons from design decisions made in another software. E.g., in OpenToonz 1316, a participant argued against the dark theme: ``\textit{I also hear nothing but complaints from people who use After Effects all the time, that it's too dark when working with particles or when the sun is shining through the window...}''

\subsection{Commitment}
Thread participants also discussed their wishes or plans as to who and how to contribute in fixing the issue.

\vspace{-4px}
\textbf{Encourage others to contribute} (\textit{n}=35). E.g., in Atom 307, ``\textit{If you want this to happen faster, perhaps you could also give that a try and post a report here or somewhere else.}'' And in OpenToonz 102, ``\textit{Would it be possible for some programmer to expose this functionality to the gui?}''

\vspace{-4px}
\textbf{Offer to contribute or collaborate} (\textit{n}=13). E.g., in Atom 5344, a participant offered to fix the issue by posting: ``\textit{I'm looking for an entry point to contribute to Atom, and seeing this as is a beginner issue, thought I could work on this?}''

\vspace{-4px}
\textbf{Need to prioritize} (\textit{n}=12). E.g., in OpenToonz 102, after a lengthy discussion on the optimal design, one participant wrote: ``\textit{I think that we should focus on bringing the broken functionality back first, then improve its presentation.}''

\vspace{-4px}
\textbf{Plan to fix soon} (\textit{n}=10). E.g., in Atom 307, one developer responded to participants' urges to support big files by posting: ``\textit{Just to be clear, improving performance and supporting larger files are ... both part of our 1.0 focus.}''

\input{margin_p5}

\subsection{Fix/Implementation}
After one or more developers started to actively fix the problem, the thread participants also discussed progress of the fix and details of the implementation.

\vspace{-4px}
\textbf{Report progress} (\textit{n}=48). E.g., in Atom 307: ``\textit{@maxbrunsfeld and I made a bunch more progress on fixing this today. It's a huge change because we're basically overhauling the entire document model... It's going to be at least a few weeks until it's ready though.}''

\vspace{-4px}
\textbf{Inquiry progress} (\textit{n}=9). E.g., in Atom 1772: ``\textit{So are there any estimations when this feature (bugfix) will become available?}'' And in OpenToonz 417 ``\textit{Any news on when we might be able to build this with Windows locally?}''

\vspace{-4px}
\textbf{Provide support for implementation} (\textit{n}=14). E.g., in OpenToonz 1316, after the developer wrote he or she was new to SVG images, one participant offered help by describing basic concepts of SVG images.

\vspace{-4px}
\textbf{Request test/feedback on prototype} (\textit{n}=15). E.g., in Eclipse Che 3614, one developer wrote, ``\textit{we have published a specification for a better Git integration in the next version of Che. Please review the issue and share your feedbacks.}''

\section{Community Differences}
While the analyzed issues have all covered the aforementioned themes, we found that those themes demonstrated different distributions across the three projects. Figure \ref{fig:IssueDistribution} shows the percentage of our coding in the four high-level categories for each project. Particularly, more than half of the discussions from the Atom issues fell into the \textit{Current State} category, while in the Eclipse Che and OpenToonz issues, this category only covered around 1/6 of the discussions. In contrast, the Eclipse Che and OpenToonz community focused more on the \textit{Desired State} of those issues. The OpenToonz community also spent considerably more space discussing fixing progress and implementation details, as well as commitment affairs.

Looking into the details in the discussion threads, we found that the \textit{Current State} discussions in the Atom community is mostly focused on debating the urgency and/or severity of the issues; i.e. the users were trying to convince the developers that the issue needed to be addressed. Those issues were eventually fixed because the community reached a consensus that they are severe problems that can greatly affect usability/UX and user adoption. The Eclipse Che community, on the other hand, put a lot of emphasis on identifying the optimal design. Its issues were usually reported by the software developers, who often also included a tentative design in order to seek feedback from the community. Those issues were fixed because the community reached a consensus on the desired state. The OpenToonz community also focused on addressing the solution space. Because this community is comprised of a lot of artists and designers, the thread participants frequently used UI mock-ups to demonstrate their ideas. They also put particular emphasis on the design details (e.g. color, icon size, spacing among UI components); this emphasis is not seen in the other two communities. However, the OpenToonz community faced difficulties looking for appropriate developers to fix the issues.

%% file: margin_p5.tex
\begin{marginfigure}[-21pc]
  \vspace{-490pt}
  \begin{minipage}{\marginparwidth}
    \centering
    \includegraphics[width=0.73\marginparwidth]{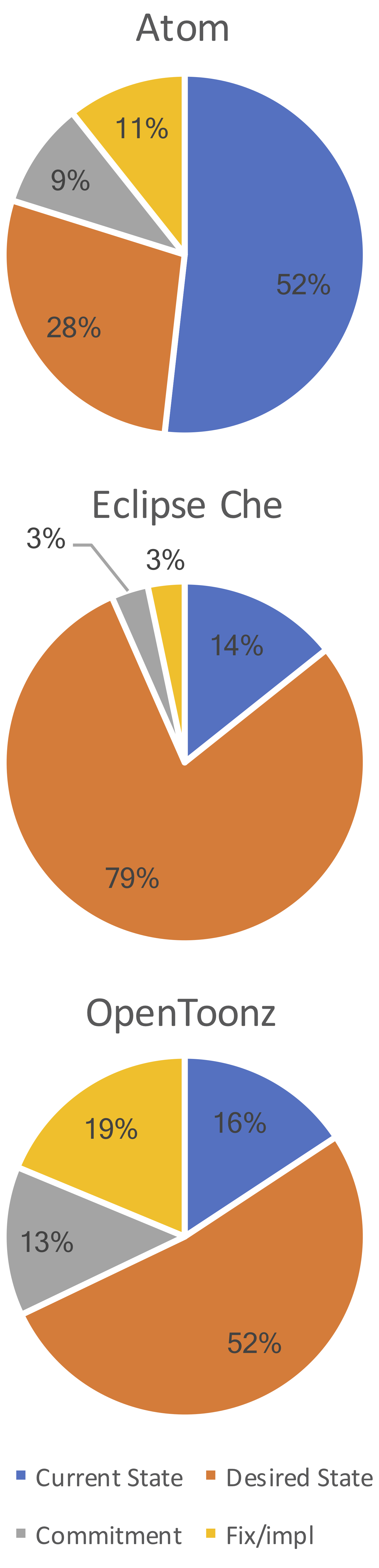}
    \caption{Distribution of themes in different projects.}
    \label{fig:IssueDistribution}
  \end{minipage}
\end{marginfigure}

%% file: s_discussions.tex
\section{Discussions}
In this paper, we explored how the OSS communities identified, discussed, negotiated, and eventually fixed usability and UX issues. While the scope is limited to only three issues from three OSS projects, this work serves as a needed first step to explore opportunities in supporting the OSS communities better manage usability and UX issues. 

In the issue threads we analyzed, participants from different projects demonstrated different emphases. The characteristics of the community may have had greatly affected the focus and style of its discussion. We hypothesize that projects for diverse users (e.g. Atom) tend to have a greater focus on the \textit{Current State} compared to projects for a more homogeneous audience (e.g. programmers for Eclipse Che and animators for OpenToonz). 
Other factors such as the nature of the issue and the stage of the project may also have had impacts.

Further, we found a common characteristic in these issue discussions: they were largely based on personal opinions and experiences, with a lot of over-generalized assumptions. As a result, lengthy debates on both the problem and solution spaces manifested in the threads.
These findings all highlighted the needs of supporting the OSS communities in managing complex issues such as usability and UX.

In the future, we plan to expand this study by (1) analyzing more threads that cover different issue types (i.e. bugs, enhancements, new features, etc.), (2) including more projects that exhibit diverse characteristics, and (3) interviewing and surveying OSS developers and users to identify primary personas. This initial work also points to important next steps that include exploring methods and tools to support diverse OSS communities (e.g. homogeneous vs. heterogeneous) in addressing usability and UX issues.